\documentclass[conference]{IEEEtran}
\IEEEoverridecommandlockouts
\usepackage{cite}
\usepackage{amsmath,amssymb,amsfonts}
\usepackage{algorithmic}
\usepackage{graphicx}
\usepackage{stfloats}
\usepackage{textcomp}
\usepackage{xcolor}
\usepackage{multirow}
\def\BibTeX{{\rm B\kern-.05em{\sc i\kern-.025em b}\kern-.08em
        T\kern-.1667em\lower.7ex\hbox{E}\kern-.125emX}}

\usepackage{booktabs} 

\usepackage{listings}
\usepackage{color}
\usepackage{verbatim}
\usepackage{ulem} 

\usepackage{etoolbox}
\makeatletter
\patchcmd{\@makecaption}
{\scshape}
{}
{}
{}
\makeatother

\definecolor{dkgreen}{rgb}{0,0.6,0}
\definecolor{gray}{rgb}{0.5,0.5,0.5}
\definecolor{mauve}{rgb}{0.58,0,0.82}

\begin{document}
    
    \title{Stage Lookup: Accelerating Path Lookup using Directory Shortcuts}
    
    \author{\IEEEauthorblockN{
            Yanliang Zou\IEEEauthorrefmark{2}\IEEEauthorrefmark{3}\IEEEauthorrefmark{4},
            Tongliang Deng\IEEEauthorrefmark{2},
            Jian Zhang\IEEEauthorrefmark{2},
            Chen Chen\IEEEauthorrefmark{2},
            Shu Yin\IEEEauthorrefmark{1}\IEEEauthorrefmark{2}
        }
        \IEEEauthorblockA{
            Email: \{zouyl, dengtl, zhangjian, chenchen, yinshu\}@shanghaitech.edu.cn\\
            \IEEEauthorrefmark{2}School of Information Science and Technology, ShanghaiTech
University, China\\
            \IEEEauthorrefmark{3}Shanghai Institute of Microsystem and Information Technology,
Chinese Academy of Sciences, China\\
            \IEEEauthorrefmark{4}University of Chinese Academy of Sciences, China}\\
    }
    \maketitle
\thispagestyle{plain}
\pagestyle{plain}
    \begin{abstract}
    
    The lookup procedure in Linux costs a significant portion of file accessing time as the virtual file system (VFS) traverses the file path components one after another. The lookup procedure becomes more time consuming when applications frequently access files, especially those with small sizes. We propose Stage Lookup, which dynamically caches popular directories to speed up lookup procedures and further reduce file accessing latency. The core of Stage Lookup is to cache popular dentries as shortcuts, so that path walks do not bother to traverse directory trees from the root. Furthermore, Stage Lookup enriches backward path walks as it treats the directory tree in a VFS as an undirected map. We implement a Stage Lookup prototype and integrate it into Linux Kernel v3.14. Our extensive performance evaluation studies show that Stage Lookup offers up to 46.9\% performance gain compared to ordinary path lookup schemes. Furthermore, Stage Lookup shows smaller performance overheads in \texttt{rename} and \texttt{chmod} operations compared to the original method of the kernel.
    \end{abstract}
    
        
    \section{Introduction}
	
\begin{figure}[htpb]
		\centering
		\includegraphics[width=0.45\textwidth]{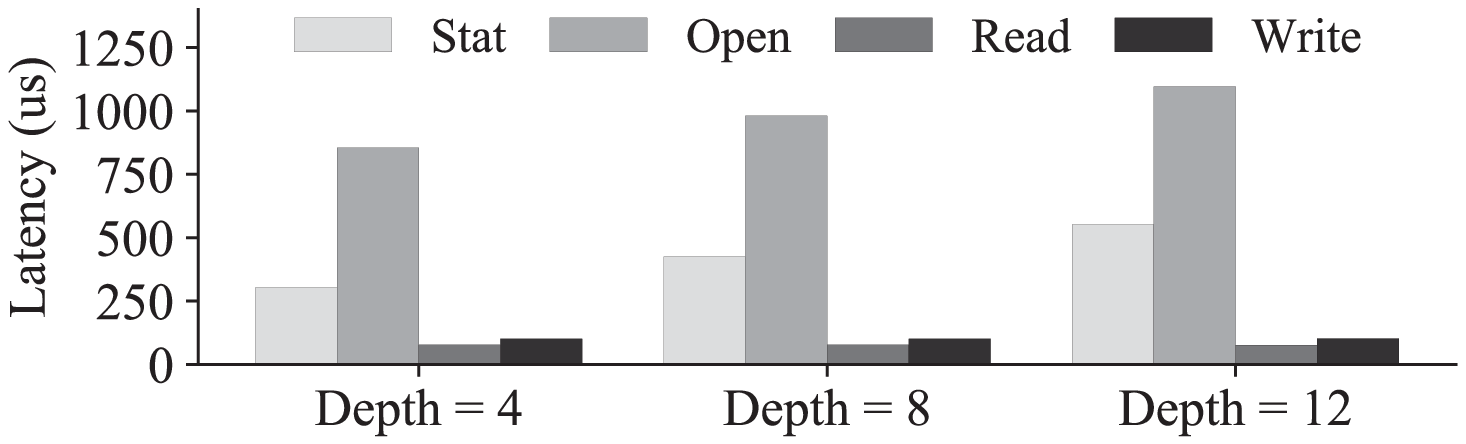}
		\caption{Latency of system calls including \texttt{stat}, \texttt{open}, \texttt{read}, and \texttt{write}}
		\label{figure:case_study}
	\end{figure}

	Path lookup is an essential procedure in operating systems because many system calls must operate on file paths at first before they can manipulate files. iBench system shows that about 10-20\% of entire system calls perform a path lookup~\cite{Harter2011AFI}. The directory cache work by Tsai \textit{et al.} presents that the path-based syscalls weight up to 60\% of execution time in many file management commands such as \texttt{find}, \texttt{tar}, \texttt{du}, and \texttt{git-diff}. These operations are mainly composed of \texttt{open} and \texttt{stat} operations (e.g., \texttt{open} weights more than 2/3 of the \texttt{find} execution time, and \texttt{stat} dominates the \texttt{git-diff} execution time)~\cite{Tsai2015}. We use LMBench\cite{lmbench} to evaluate latency of four basic file-related syscalls including \texttt{stat}, \texttt{open/close}, \texttt{read}, and \texttt{write} (see Fig.~\ref{figure:case_study}). The results show that the path lookup operation occupy the main latency of file managements.
	
	Not only does the path lookup consume a large fraction of execution time, but it is of frequent occurrence in file operations. Smartphone applications such as File Explorer and NoxCleaner produce more than 10,000 path lookups per seconds\cite{shao2019}. We analyze operational statistics from the TaihuLight supercomputer and find that up to 52.4\% operations require path lookups in a day\cite{Yang2019,Ji2019}. Further evaluations of a one-day statistic show that 13 million path lookup requests generate more than 89 million dentry searches, while only 235 distinct dentries are involved. In other words, only 14.6\% of dentry searches are effective and a great number of dentries are repeatedly visited.

	Improving the efficiency of path lookup brings a fast response of applications~\cite{Daeho2015}. Studies show efforts to improve the path lookup latency at the file system level\cite{Shuanglong2016,LocoFS}. For example, Ren \textit{et al.} designs TableFS to translate the local file system as an object-store\cite{TableFS}. In the parallel file system and distributed file system, researchers prefer to reduce latency by optimizing metadata management\cite{SoMeta,DeltaFS}. 
	
	Other studies try to optimize the path lookup efficiency at the dentry level. Tsai \textit{et al.} purposes a full-path directory cache mechanism to reduce the lookup latency by storing recently accessed dentries that are hashed by full and canonicalized paths~\cite{Tsai2015}. However, this mechanism introduces an extensive overhead to maintain the path components hash table when a directory needs to be modified via commands such as \texttt{rename} or \texttt{chmod}. Other than caching the recently accessed dentries, Han \textit{et al.} proposes a mechanism to cache frequently accessed prefixes to improve path lookup performance on smartphones\cite{shao2019}. But applications with large directory trees may suffer from low prefix hit rate and poor path lookup efficiency due to the limited cache size. Directory modifications including renames and change modes are also challenging on mobile devices.
	
	Besides, the path lookup optimization needs to take file access patterns into account for two reasons:(1). files that are associated with the same application tend to be accessed together as they are commonly stored under the same directory\cite{Xiaoning2007,song2013}; and (2). small files are accessed more frequently\cite{Harter2011AFI,Shuanglong2016}. 
	
	We propose a method called \textit{Stage-Lookup} to accelerate path lookup procedures and reduce file lookup latency in VFS. Stage-Lookup is designed for a common scenario that files are stored in a directory tree with more than one level (i.e. files are not stored directly under the root directory). By caching frequently accessed nodes in a directory, Stage-Lookup can reduce the number of path components for path-based syscalls, hence reduce the lookup latency. We call these nodes \textit{pivots} and maintain a \textit{Pivot Pool} in the memory to cache them based on dentries' accessing frequency. Notice that the path between two specific nodes in a directory tree is unique. Every time a path-based request arrives, the system kernel chooses the optimal pivot of the given path in Pivot Pool and raises conventional operations starting from the chosen pivot. If the pivot is not included in the path, Stage-Lookup must roll up levels to one of its ancestor directories, which is also a path component of the given path. This operation can be skipped if the ancestor dentry is marked (details are discussed in Section~\ref{subsection:finding_pivots}). 
	
	The major challenge of this research is to efficiently find the best pivot and management of Pivot Pool. We apply an ascending-ordered list to maintain and search suitable pivots. Besides, Stage-Lookup performs string comparison only once for the given path instead of complex dentry searches. Moreover, Stage-Lookup can run permission checks quickly as every pivot stores its ancestors' permission information, any lookup procedure will carry out a permission check on the pivot right away. Stage-Lookup also shows its advantages in handling directory modifications especially renames and change modes because these metadata operations only involve a limited number of pivots. 
	
    The main contributions of this paper include:
    \begin{itemize}
    	\item We propose a path lookup optimization scheme in the VFS layer called Stage Lookup, which comprises techniques including:
    	\begin{itemize}
    		\item Two-Stage path lookup: picking an optimal pivot as the start point, an then walking to the target from the pivot;
    		\item Pivot pool management: generating new pivots and updating pivots to ensure the two-stage path lookup procedure can always choose the optimal pivots;
    		\item Directory metadata modification: let pivots to inherit their ancestors' directory metadata to restrain the modification procedure in the scope of pivots.
    	\end{itemize}
    	\item We implement Stage Lookup on Linux kernel v3.14 and provide a comprehensive experimental study  to fully evaluate the efficacy of Stage Lookup with microbenchmark and real-world workloads.
    \end{itemize}
    
    \section{Background}
	\subsection{Original Lookup}
	When the Linux kernel tries to lookup a file with a given path, it parses the path into components and traverses them one by one. The lookup procedure accesses each component's metadata for permission checks. And the kernel will access storage devices to get the required path components if they are not cached in the main memory.
	
	Linux kernel builds a structure called dentry for each directory and caches them in the memory to avoid duplicated efforts of accessing file systems for certain directories\cite{BovetBook}. The kernel maintains a hash table for dentries searches. Dentries with the same hashing value are aggregated into a bucket in the hash table. The lookup process parses the first component from the path and calculates its hash value to target the corresponding bucket. Then it scans dentries in the bucket and determines the target ones by checking their ancestors and comparing the name strings. If the subsequent component does not exist in the directory cache, the process will collect the component's inode ID from the current directory and then access the inode to build a corresponding dentry. 
	
	\subsection{Full-path Indexing}
	Full-path indexing is commonly studied to optimize the performance of indexing a file at the file system level\cite{fullpath_indexing, direct_looup} and at the dentry level\cite{Tsai2015}. 
	
	As a state-of-art study, Tsai's propose a directory cache design that modifies the structure of dentries. It maintains a hash table to map a full-path to a specific dentry. When indexing a file, the kernel will calculate a hash value for the full path and look up its corresponding dentry in the hash table Tsai's design also maintains a cache for the permission check. Every time a new dentry's permission is checked, an entry containing the permission to this dentry will be inserted into the cache.
	
	If the entry of a given path for indexing does not exist in the cache or out-of-date, the kernel has to switch back to the original path lookup procedure. In this case, the permission checks become noticeable overhead. This directory cache design works well when a directory has been accessed before (i.e., the directory is cached in the main memory) or users do not frequently change permissions and names of files. When a file changes its permission or name, not only the indexing efficiency is affected but it introduces more dentry traversals to update the relevant entries. A quick test indicates that Constant-latency operations(i.e., \texttt{chmod} and \texttt{rename}) in the original lookup may become linear in the size of Tsai's directory cache, and take up to 100 times of latency to complete such an operation.
	
	Besides, how to efficiently handle directory metadata modification is a common problem for the full-path indexing strategy. For example, BetrFS is local file system based on $B^{\epsilon}$-$tree$ , a structure to realize full-path indexing\cite{BetrFS}. BetrFS proposes a coupling optimization to particularly handle this problem\cite{fullpath_indexing}. BetrFS uses a mechanism called Tree Surgery to carve up nodes on the $B^{\epsilon}$-$tree$, in order to ensure consistency and efficiency when \texttt{rename} and \texttt{chmod} occur. 
	
	However, in the VFS layer, the performance of full-path indexing still suffers from handling metadata modification. This motivates us to propose a path lookup optimization scheme to optimize both the lookup latency and the directory metadata modifications including \texttt{rename} and \texttt{chmod}. 
	
	\section{Design of Stage Lookup}
	\label{section:design}
	\subsection{Overview}
	\label{subsection:overview}
	
    We can observe from the investigation on operational statistics of the TaihuLight supercomputer that different path walks commonly share some prefixes with each other. Path lookups will be much more efficient if we can cache the commonly accessed prefixes and start path walks from them instead of starting all the way back from the root (``/").
    
\begin{figure}[htpb]
		\centering
		\includegraphics[width=0.45\textwidth]{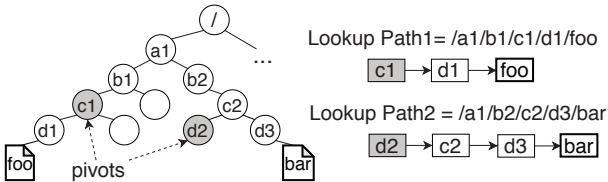}
		\caption{A Stage Lookup path walk example. Grey nodes represent pivots.} 
		\label{figure:stage_lookup}
	\end{figure}

    The design goal of Stage lookup is to reduce the path lookup latency by minimizing the number of dentries visited during the operation. Stage Lookup introduces a pivot structure, which stores the commonly accessed dentries so that most path lookups can start at an optimal pivot instead of the root (see Fig.\ref{figure:stage_lookup}, case ``Path1"). In order to find the file \texttt{foo}, Stage Lookup picks the pivot \texttt{c1} and starts the path walk from \texttt{c1} to its descendant directory (\texttt{d1}), then reaches to the target file \texttt{foo}. Compared to the original path lookup that starts from the root directory (``/"), Stage Lookup only walks two path components. Furthermore, Stage Lookup supports backward pivot walks, meaning that path walks can roll up one level to the chosen pivot's ancestor directory then walks down to another descendant directory (see Fig.\ref{figure:stage_lookup}, case ``Path2"). Stage Lookup starts at the pivot \texttt{d2} to look for the file \texttt{bar}, even though \texttt{d2} is not a direct ancestor directory node to \texttt{bar}. Stage Lookup first rolls up to \texttt{d2}'s ancestor \texttt{c2} and then walks down to \texttt{d3} before it finds \texttt{bar}. In this case, Stage Lookup still walks fewer components than what the original lookup operation does.
    
\begin{figure}[htpb]
    	\centering
    	\includegraphics[width=0.45\textwidth]{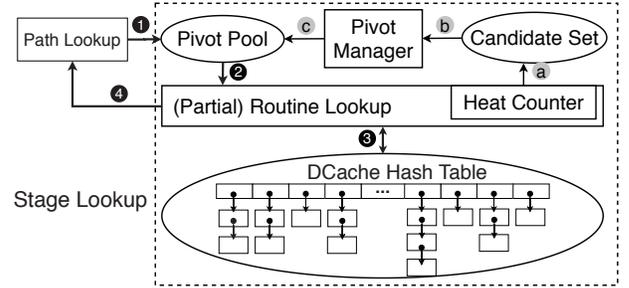}
    	\caption{The architecture and workflow of Stage Lookup. The Stage Lookup workflow consists of four steps: (1) look for the optimal pivot in Pivot Pool with the given path (Stage One); (2) pick the optimal pivot as the path lookup starting point; (3) walk the path components in the DCache Hash table to search the target dentry; and (4) return the dentry and finish the path lookup. Stage Lookup manages pivots in three steps: (a) inserts frequently access directory nodes to Candidate Set as pivot candidates; (b) Pivot Manager determines pivots from Candidate Set; and (c) new pivots are pushed into Pivot Pool.}
    	\label{figure:architecture}
    \end{figure}

    We then discuss the design details of Stage Lookup. Fig.\ref{figure:architecture} shows the architecture and workflow of Stage Lookup. Stage Lookup is comprised of two modules (Heat Counter and Pivot Manager) and two data structures (Candidate Set and Pivot Pool):
   
   \textbf{Heat Counter} calculates the popularity of a directory node with a heat value. The heat value is self-incremented by one when the node is accessed. The higher the heat value of a node, the more frequently the node is accessed, and more than likely such node will be a pivot candidate. We explain the heat value updating strategy in Section~\ref{time-window-heat}.
   
   \textbf{Candidate Set} stores directory nodes with high heat values. These nodes are treated as pivot candidates for Pivot Manager that determines which candidates are pushed to Pivot Pool. Candidates will be evicted from Candidate Set if their heat values are surpassed by others, indicating the access frequency of other nodes is increasing.
   
   \textbf{Pivot Pool} accommodates all the pivots for Stage Lookup. Each pivot contains descriptive information of a frequently accessed directory. Most pivots in Pivot Pool stay statically before Pivot Manager updates the pool. But pivots have to be updated immediately to retain consistency if directory metadata modification operations (i.e. \texttt{rename} and \texttt{chmod}) are performed.
   
   \textbf{Pivot Manager} updates pivots in Pivot Pool by generating new pivots from Candidate Set and recycling old pivots in the next period. This manager is running in the background and is awakened periodically.
    
    Stage Lookup performs two stages for a path lookup operation:
    
    \textbf{Stage One: } Find a pivot which is much closer to the target than the root (``/"), and pick the chosen pivot as the starting point of the path lookup. (Steps (1)\&(2) in Fig.\ref{figure:architecture}). 
    
    \textbf{Stage Two:} Walk the path components in the DCache Hash Table to search the target dentry, then return the dentry to finish the path lookup operation. (Steps (3)\&(4) in Fig.\ref{figure:architecture}). The cache hash table is an intrinsic container to maintain dentries in the Linux kernel.
	
	\subsection{Finding An Optimal Pivots}
	\label{subsection:finding_pivots}
	
	Performing a path lookup from a directory node (a.k.a. pivot) that is much closer to the target than that from the root (``/") is the major advantage of the Stage Lookup mechanism. How to quickly find the optimal pivot becomes the main challenge of Stage Lookup.
	
	A simple way is to compare each pivot's path with the given path and find the optimal one. Although this method makes it easy to manage pivots in Pivot Pool, the time complexity to find the best-fit pivots is too high to be acceptable. To solve this problem, we arrange the pivots as a list in ascending order of their paths. Each list entry stores the length of the shared prefix so as to reduce the repetitive string comparison of a path.
	
	\begin{figure}[htpb]
		\centering
		\includegraphics[width=0.45\textwidth]{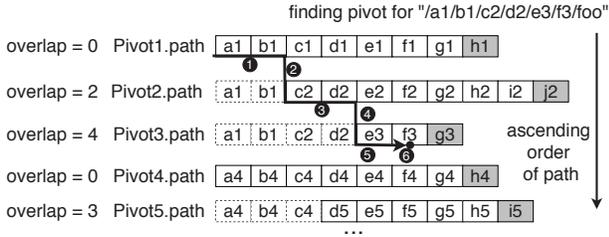}
		\caption{The procedure of finding the best pivot in Pivot Pool. Each item represents a component of the pivot's path and the grey color marks the pivot. \texttt{Overlap} gives the number of components that a pivot's path shares with the previous one. Step 1-6 shows an example: a) the process compares the given path with Pivot1 (Step 1); b) it meets a distinct component and finds that Pivot2 shares the same prefix by checking its \texttt{overlap} (Step 2); c) Step 3-5 act similar; d) When the process find that Pivot4 does not share the same prefix with the given path, it stops(Step 6) and Pivot3 is the best pivot.	}
		\label{figure:finding-pivot}
	\end{figure}
	
	Fig.~\ref{figure:finding-pivot} presents a Stage Lookup example of finding an optimal pivot to target file \texttt{foo} at \texttt{/a1/b1/c2/d2/e3/f3/foo}. Stage Lookup firstly compares the given path with pivot1 (\texttt{Pivot1.path}) then turns to compare with pivot2 (\texttt{Pivot2.path}) from its third component because the path does not match with the third component of pivot1 Stage Lookup then jumps to pivot3 (\texttt{Pivot3.path}) after it finds out the path does not match with pivot2's fifth component and finalizes the comparison in pivot3 when it finds \texttt{f3}. Pivot3 will be picked as the optimal pivot even though it caches the directory node \texttt{g3} which is one level lower than the requested one (\texttt{f3}). Notice that Stage Lookup may not find a pivot that perfectly matches the given path, but it can find a pivot which is closer enough to the path so that the path lookup procedure can be much faster than that from the root (``/").
	
	In order to find \texttt{foo}, Stage Lookup then processes Stage Two, which performs the path lookup from pivot3 (\texttt{g3}). Stage Lookup first rolls one level up to pivot3's ancestor node (\texttt{f3}) and succeeds in finding \texttt{foo}.
	
	In order to minimize the overhead of a directory rolling upwards, we keep the dentry pointer to each component of the chosen pivot. In this case, the lookup procedure can directly access the valid dentry, neglecting the distance between the chosen pivot (\texttt{g3}) and the valid component (\texttt{f3}).
	
\begin{figure}[htpb]
		\centering
		\includegraphics[width=0.45\textwidth]{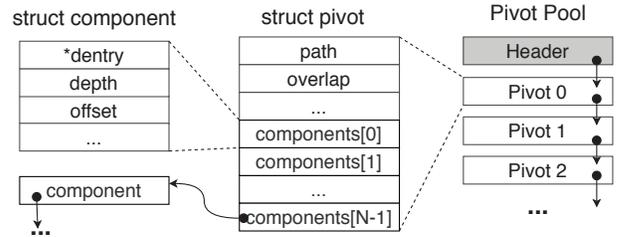}
		\caption{Pivot Pool data structure. Each pivot in the pool is comprised of its path, overlap value, and components. The number of components is set to N by default, but it can be extended with a list pointer.}
		\label{figure:data-structure}
	\end{figure}

	Fig.~\ref{figure:data-structure} demonstrates the data structure of a pivot and its components. Pivot Pool is a list of \texttt{struct pivot} in ascending order of their path. The header of Pivot Pool stores a pointer that points to the first pivot. The \texttt{struct pivot} is comprised of pivot's path, overlap value, and an fixed-sized array of \texttt{struct component}. The \texttt{overlap} stores the number of shared path components between the current pivot with the pivot that sits above it in the Pivot Pool list. The \texttt{struct component} includes a \texttt{struct dentry} pointer, depth, and offset. The fixed-size array may waste space for short paths, but it can diminish data loading operations from memory to CPU cache. We keep a list pointer for \texttt{struct pivot} to extend the extra components to adapt the pivot structure to the path's arbitrary length.
	
    \subsection{Maintaining Pivots}
    \label{subsection:pivots}
    
    After explaining the way to find optimal pivots, we then describe how to pick and update pivots in this section.
    
    \subsubsection{Heat Value Updating Strategy}
    \label{time-window-heat}
    As aforementioned, frequently accessed directory nodes will be considered as pivots. Stage Lookup uses Heat Counter to set a heat value for dentries, representing the access frequency of lookup targets.
    
    Heat Counter increases the target's heat value by one when executing a path lookup. We do not increase the heat values of other involved path components because it will eventually make the top nodes of the directory tree hold the highest value. Besides, a pivot stores dentries of all the involved nodes so that Stage Lookup can traverse to any ancestor nodes of the pivot. We do not adopt dynamic weight to different directory depth as it will introduce path lookup overhead. Because a process has to scan the path to collect the weight information before it executes a path lookup.
    
    However, the heat values can not be valid all the time. For example, if a directory lies idle for a long time, it does not deserve a pivot although it used to be required frequently. 
    
    Thus, we set a validity period for all the heat values, which is identified by a version number. We further set a global version number managed by Pivot Manager. A dentry's heat value is valid only if its version number matches the unique global one; otherwise, it will be reset when new access occurs. The global version number is updated by Pivot Manager only when a new period comes. The length of the period is static in our study, which could be dynamic in our future work. But it can be adjusted by a system call with any other value. 
    
    \subsubsection{Candidates for pivots}
    \label{subsubsection:candidate}
    
    At the end of each period, Pivot Manager will be waked up to update pivots. However, it is inefficient to traverse the whole DCache to find out a number of dentries with outstanding heat values. According to the statistic of a study\cite{direct_looup}, about one million directories are contained in a local file system.
    
\begin{figure}[htpb]
		\centering
		\includegraphics[width=0.45\textwidth]{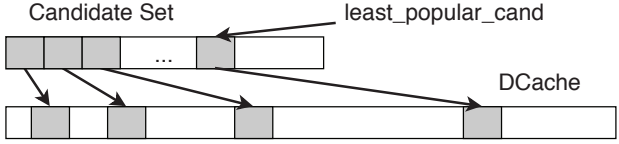}
		\caption{Candidate set collects the most popular dentries in the current period. And \texttt{least\_popular\_cand} points to the item that may be replaced by a new member.}
		\label{figure:candidate}
	\end{figure}

    Thus, We introduce Candidate Set to gather a number of popular dentries as candidates of pivots. We set a pointer called \texttt{least\_popular\_cand} to mark the item whose heat value is ``least" frequently updated in Candidate Set (see Fig.\ref{figure:candidate}). During a lookup operation, the process compares the target's heat value with the sum of a threshold and \texttt{least\_popular\_cand}'s heat value. If the former wins, the target's dentry will replace what \texttt{least\_popular\_cand} points to as a candidate and inherit the pointer. The threshold is to prevent some popular dentries from being in and out frequently in a short time.
    
    
    Besides the impact of a new member, the \texttt{least\_popular\_cand} can also be updated by candidates internally. Everytime a candidate's heat value is updated, it will compare its value with the \texttt{least\_popular\_cand}'s, and the loser becomes the new \texttt{least\_popular\_cand}. Thus, \texttt{least\_popular\_cand} does not strictly mean the candidate with the least heat value in the set. For example, if a dentry is never accessed after being a candidate, it can prevent itself from being the \texttt{least\_popular\_cand} because no heat value updating means no comparison with the \texttt{least\_popular\_cand}.
    Maintaining \texttt{least\_popular\_cand} offers an effective rather than the best strategy to update the set, since traversing the whole set to find out the real least-accessed candidate is costly. We believe that every candidate in the set will be accessed frequently.    
    
    When a new period comes, Pivot Manager will generate new pivots with dentries serving by Candidate Set. For each dentry, it will traverse all its ancestors to get the whole path.

    \subsection{Handling Metadata Modification}
    Besides improving the performance of path lookup, Stage Lookup can efficiently handle the metadata modification of directory, such as \texttt{rename} and \texttt{chmod}.
    
    Modifying directory's metadata, such as \texttt{chmod}, \texttt{rename}, \texttt{move} and etc., is always a potential problem for full-path indexing methods. Because these methods use the whole path string as a key to attain hashing value, which means they have to depend on extra mechanisms to guarantee permission check and directory operations. It may introduce considerable overheads.
    
    In Stage Lookup, modifying directory's metadata only affects Stage One while Stage Two follows the conventional lookup procedure. Considering that a common component may be included in multiple pivots, we can directly remove and free those related pivots from Pivot Pool before executing the modification.
    Since the size of Pivot Pool is limited, we can say that our overheads to these problems are in constant time complexity.
    
    
    \subsection{Concurrency}
    \label{subsection:concurrency}
    
    Besides the modification of metadata, another important issue for Stage Lookup is the concurrency problem, which can be considered respectively in two situations: updating Pivot Pool and modifying metadata.
    
    The first one occurs when updating Pivot Pool periodically. We have to consider concurrency among the working processes. However, it is not an efficient way to block the whole Pivot Pool to insert new items and free out-of-date items.
    
    Thus, we set up another pool as a collaborator. The two pools act as a working pool and a waiting pool by turns. When a new period comes, Pivot Manager generates new pivots in the waiting pool, say \texttt{Pool B}. The working one, say \texttt{Pool A}, keeps unaffected until Pivot Manager finishes generating. Then the \texttt{Pool A} will be totally replaced by the fresh \texttt{Pool B} if no references fall on it. We utilize \textit{RCU(Read Copy Update)} to achieve efficient monitoring of Pivot Pool's references, which is a technique to handle the concurrency in the kernel. With \textit{RCU}, new requirements for pivots will be guided to the \texttt{Pool B} if it is ready, while the current working processes can still stay at the \texttt{Pool A}. When all cores of the machine complete at least one context switch respectively, it can be confirmed that no processes are accessing the \texttt{Pool A}. Because in the kernel, the path lookup procedure is wrapped by \textit{rcu\_read\_lock()} and \textit{rcu\_read\_unlock()} which prevent the process from context switch before finishing. After the two pools achieve such an exchange, Pivot Manager will clean up the old \texttt{Pool A}.
    
   Pivot Manager will upgrade the global version number after generating pivots, which means a new round comes for all the dentries to join Candidate Set. Then Pivot Manager will kick those overdue members out of Candidate Set after exchanging the two Pivot Pool.  
   
   For the second situation, modifying a directory's metadata(\texttt{rename}/\texttt{chmod}) causes a consistency problem since a pivot is related to its two neighbors in the ascending-order list.
   Supposed Path \texttt{P\_a} describes the target being modified, and \texttt{p\_v} is the first pivot contains this target in the working pool(\texttt{Pool A}). We set a flag on \texttt{p\_v} to mark that \texttt{p\_v} and all the following pivots are temporarily invalid for new accesses. The other pivots lying before \texttt{p\_v} is unaffected. Then we remove those pivots related to \texttt{P\_a}, and reactivate the rest pivots after that. However, Pivot Manager may be generating new pivots covering \texttt{P\_a} in the alternate pool, say \texttt{Pool B}. In order to reduce the latency of \texttt{rename}/\texttt{chmod}, we mark \texttt{Pool B} as totally invalid since it is not in used. So that the \texttt{Pool B} will be cleaned but no pool exchanging occurs in the coming period, which means that \texttt{Pool A} will be still valid in the next period.

    \subsection{Cost Analysis}
    \label{subsection:cost-analysis}
    Stage Lookup brings convenience to the path-lookup procedure. Here we analyze the costs of different aspects of Stage Lookup.
    
    \textbf{Time Complexity of Finding Pivot}:
    When executing a path lookup, the original method will parse the string into components. Each component string will be scan twice: one is to calculate the hash value; the other is to verify the dentry's name. Similarly, full-path indexing takes two scans on the whole path for the same reasons.
    
    Differently, stage lookup will not complete the dual scans. Let $p=p_1+p_2$ be a given path where $p_1$ represents the prefix skipped in Stage One, and $p_2$ be the rest path used in Stage Two. \texttt{Overlap}(see Section~\ref{figure:finding-pivot}) helps Stage One to take only one scan to work out $p_1$. And Stage Two is similar to the original lookup procedure, which will traverse the rest path twice.
    
    \textbf{Overheads of Removing Pivots}: 
    We discussed how to maintain Pivot Pool when metadata modification occurs in Section~\ref{subsection:concurrency}. It is to make partial pivots invalid and remove those related pivots. However, freeing a data structure takes time which may lead to an unacceptable latency for the whole operation. Fortunately, \textit{RCU} mechanism serves an asynchronous way to solve this problem. By inserting the removal into a callback function which will be detected and executed by the \textit{RCU soft interrupt} asynchronously, the current process can directly move on without waiting. Thus, metadata modification can endure little overheads.
    
    \textbf{Space Overheads}: 
    To achieve the candidates and the heat value management, we create four new members for \texttt{struct dentry}, which worth 30 bytes in a 64-bit system or 18 bytes in a 32-bit system. Instead of simply inserting new members into \texttt{struct dentry}, we occupy the partial space of \texttt{d\_iname} which is a \texttt{char} array to keep \texttt{struct dentry} aligned on 64 byte cachelines. Reducing the space of \texttt{d\_iname} will not affect the function of \texttt{struct dentry}. So that we introduce new members in \texttt{struct dentry} without breaking its alignment on 64 bytes cachelines. As a comparison, Tsai's full-path indexing method\cite{Tsai2015} adds an extra 88-bytes overheads for each dentry structure to satisfy their requirement, which breaks the alignment on cachelines.
    
    In Candidate Set, the members are dentries linked to each other as a list through two internal pointers which is included in \texttt{struct dentry} as new members. Thus, candidates take no extended space in our design. 
    
    Different from a candidate, a pivot is a new data structure in the memory. Memory footprint of a pivot depends on the number of \texttt{struct component}(see Fig.\ref{figure:data-structure}). Our implementation takes about 7KB for a 16-sized Pivot Pool in the 64-bit system, suppose each pivot hold eight \texttt{struct component}.
    
    \section{Evaluation}
    \label{section:evaluation}
    
    In order to evaluate the performance of Stage Lookup, we compare Stage Lookup with the other two strategies, say the kernel's Original Lookup and Tsai's directory cache \cite{Tsai2015}.
    Since Tsai's strategy is developed on Linux kernel v3.14, we implement Stage Lookup on the same version of the kernel. 
    
    All our tests are executed on a server with 4-cored 3.3 GHz Intel Core Xeon CPU, 8GB RAM, and a 1TB 7200 RPM disk formatted as an ext4 file system. The operating system is Ubuntu 14.04 Server equipped with a 64-bit Linux kernel v3.14. Rather than overwriting the original system calls, we create new system calls such as \textit{stage\_stat}, \textit{stage\_open}, \textit{stage\_rename} and etc. in the kernel. 
    
    \subsection{Lookup performance}
    We use the v3.0 LMBench microbenchmark to measure the latency of \texttt{stat} and \texttt{open}. 
    
    \begin{figure*}[htpb]
		\centering
		\includegraphics[width=1\textwidth]{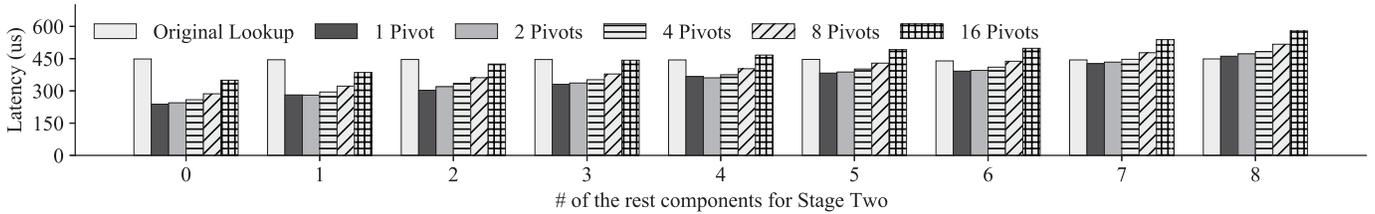}
		\caption{Performance of Stage Lookup for two cases: 1) different numbers of components for Stage Two, and 2) different numbers of pivots within Pivot Pool. All the workloads are paths with 8 components. For example, in the second group of bars, we let Original Lookup and Stage Lookup execute \texttt{stat} for the same path with 8 components. And Stage Lookup can skip the first six components through Stage One. 
		}
		\label{figure:lookup_depth}
	\end{figure*}
    
    We explore the performance of Stage Lookup when Stage Two walks different lengths. Further, we show the impact of the pivot's quantity during the searching. In Fig.\ref{figure:lookup_depth}, the nine groups of bars give the latency of Stage Lookup when Stage Two has to walk 0-8 components. The first bar in each group represents the latency of Original Lookup. And the rest bars show the performances of Stage Lookup when Pivot Pool includes different numbers(1,2,4,8,16) of pivots. The first group represents that Stage One can skip the whole given path, while the last group means that Stage One completely becomes overheads since all workloads are paths with exactly eight components. Stage Lookup performs up to 46.9\% compared to Original Lookup. 
    
    Fig.\ref{figure:lookup_depth} also shows the characteristic of Stage Lookup's overheads in. On the one hand, it is reasonable that the more components for Stage Two, the worse Stage Lookup behaves. Because it is similar to Original Lookup and affected by the path length(see Fig.\ref{figure:case_study}). On the other hand, redundant pivots we insert are totally useless for Stage One but slow down the process, while only the last pivot is effective. However, path lookup in the real-world can more probably benefit from pivots since they are those popular nodes on the directory tree.
    
    \subsection{Real-world Simulation}
    Then we turn to evaluate the performance of the three lookup strategies in a real-world environment. In a high-performance computing(HPC) cluster, I/O forwarding nodes receive requests from hundreds of computing nodes. And the parallel file system, Lustre, for example, acts as an underlying file system by installing a client on the I/O forwarding node. So that VFS-level path lookup on an I/O forwarding node is similar to a local machine in some way. We simulate an I/O forwarding node on our server and run real-world workloads on it. 
    
    \begin{figure}[htpb]
		\centering
		\includegraphics[width=0.45\textwidth]{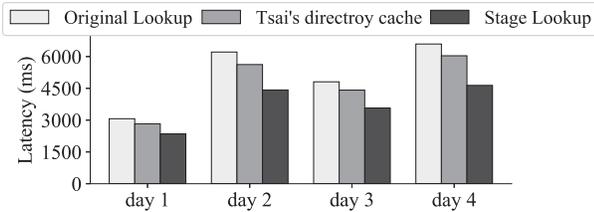}
        \caption{Latency of replaying real-world workloads which are \texttt{open} operations on an I/O forwarding node of TaihuLight Super Computer.}
		\label{figure:replay_open}
	\end{figure}

	We replay the \texttt{open} operations occurring in 4 days based on what Beacon\cite{beacon} tracked on TaihuLight Supercomputer. We generate a corresponding directory tree with random-sized(0-1MB) files as leaves. To compress the time span, we replace the large intervals between operations by four seconds, while Stage Lookup's update period is set as two seconds. 
	
	Fig.\ref{figure:replay_open} shows the comparison of Original Lookup, Stage Lookup, and Tsai's directory cache method when simulating the I/O forward node. To eliminate the impact of system cache, we clean the inode cache, dentry cache, and page cache before running every single measurement. And each result is the average of six repetitions. 
	Stage Lookup leads the performance in the comparison by showing up to 29.6\% compared to Original Lookup and 21.3\% to Tsai's design. Computing nodes in HPC are always running applications whose files are in a directory, which means that Stage Lookup can always skip various lengths of prefixes. Tsai's directory cache has to initialize each dentry that it meets first. Further, the dual scans of a given path in Tsai's strategy(see Section~\ref{subsection:cost-analysis}) slows down its performance when comparing with Stage Lookup. Thus, Stage Lookup shows the leading position in this test.
    
    \subsection{Modifying Metadata}
    Schemes for modifying a directory/file's metadata are different among the three lookup strategies. For example, when renaming a directory, Original Lookup of the kernel directly modifies the target's dentry and inode, bring little overheads to the other directories. However, before modifying, Tsai's directory cache has to visit all the target's children in the DCache to increase their version values and remove all related entries in its customized cache. Stage Lookup needs to set a flag to invalid a part of Pivot Pool; then "remove" those related pivots from the pool which is asynchronously done by the \textit{RCU} soft interrupt. 
    
    We generate a six-level directory tree owning 10,000+ nodes. Each directory in the first four levels contains 10 children, and the directories of the fifth level include a single 4KB file as a leaf for each. Before every measurement, we free all the page cache, dentry cache, and inode cache. Next, we warm up these system caches and customized cache by executing \texttt{stat} for all nodes of the whole six-level tree. 
    
    \begin{figure}[htpb]
		\centering
		\includegraphics[width=0.5\textwidth]{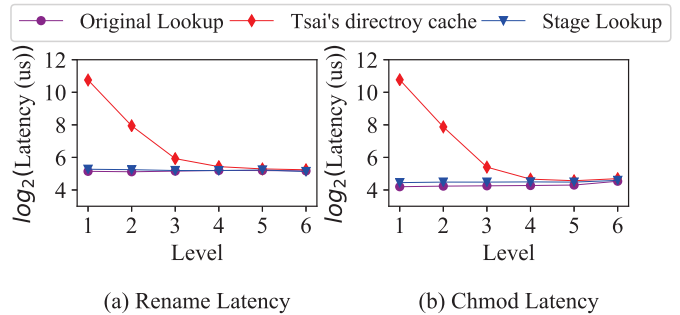}
		\caption{Performance of \texttt{rename} and \texttt{chmod} system calls}
		\label{figure:rename_chmod}
	\end{figure}
    
    Fig.\ref{figure:rename_chmod} gives the results of our comparison among the three methods. We work out our results by averaging 10 measurements' results each of which we sum up the latency of single \texttt{rename} or \texttt{chmod} operation on a file. Since Stage Lookup holds a limited-size Pivot Pool in ascending order, the process can quickly locate and modify related pivots. And the asynchronous removal of pivots saves latency for a modification. Thus, Stage Lookup introduces little overheads compared with Original Lookup. Tsai's directory cache, the full-path indexing strategy, is intimately affected by the number of directories, which is related to the target in the dentry cache. Thus, the larger its hash table is, the slower it executes \texttt{rename}/\texttt{chmod}. It may cause up to 48.8x latency for \texttt{rename} and 94.9x for \texttt{chmod} when modifying the directory in Level 1 containing 10,000+ files and directories.

    \section{Conclusion}
    This paper presents a method called Stage Lookup in the VFS level that reduces the latency of looking up a file. Rather than starting at the root("/") directory, we take one of the chosen popular directories/files named pivots as the start point which is closed to the target. Then the process will go forward or backward from the pivot to meet the target. We introduce a heat value to tell how popular a directory/file is, with which the time-window updating strategy can select a number of popular directories/files as pivots periodically. Different from some full-path indexing studies, we can quickly handle metadata modification, such as \texttt{rename} and \texttt{chmod}. We implement the Stage Lookup on the Linux kernel v3.14 and compare it with the Original lookup and a full-path indexing study. It decreases up to 46.9\% path lookup latency compared with the original lookup in the vanilla kernel; and 39.4\% compared with a state-of-art study.
    

\end{document}